\documentclass[pra,10pt,preprint,superscriptaddress,showpacs]{revtex4-1}

\usepackage{epsfig}
\usepackage{amsmath}
\usepackage{graphicx}
\usepackage{graphics}
\usepackage{float}
\usepackage{amssymb}
\usepackage[usenames,dvipsnames]{color}
\usepackage{natbib}
\usepackage{epstopdf}
\usepackage{verbatim}
\usepackage{wrapfig}

\newcommand{\bra}[1]{\left\langle #1\right|}
\newcommand{\ket}[1]{\left| #1\right\rangle}
\newcommand{\ip}[2]{\langle #1|#2\rangle}
\newcommand{\etal}{{\it et al.}}
\newcommand{\jn}[1]{~{\bf #1}}
\newcommand{\opav}[3]{\left\langle #1 | #2 | #3 \right\rangle}
\newcommand{\av}[1]{\left\langle #1 \right\rangle}

\begin{document}

\title{Amplification and suppression of system-bath correlation effects in an open many-body system}

\author{Adam Zaman Chaudhry}
\affiliation{NUS Graduate School for Integrative Sciences
and Engineering, Singapore 117597, Singapore}
\author{Jiangbin Gong}
\email{phygj@nus.edu.sg}
\affiliation{NUS Graduate School for Integrative Sciences
and Engineering, Singapore 117597, Singapore}
\affiliation{Department of Physics and Center for Computational
Science and Engineering, National University of Singapore, Singapore 117542,
Singapore}

\begin{abstract}

Understanding the rich dynamics of open quantum systems is of fundamental interest to
quantum control and quantum information processing.
By considering an open system of many identical two-level atoms interacting with a common
bath, we show that effects of system-bath correlations are amplified in a many-body system via the generation of
a short time scale inversely proportional to the number of atoms. Effects of system-bath correlations are therefore considerable
even when each individual atom interacts with the bath weakly.  We further show that
correlation-induced dynamical effects may still be suppressed via the dynamical decoupling approach, but they present a
challenge for quantum state protection as the number of atom increases.

\end{abstract}

\date{\today}
\maketitle

\section{Introduction}

The rich dynamics of open quantum systems continues to attract great interests \cite{breuerbook}. Two main motivations for now
are to achieve better quantum control and to better understand possible quantum effects in large systems.
However, the general complexity of open quantum systems puts exact solutions out of reach.  A variety of approximations or assumptions are therefore needed,
with their validity under close scrutiny in recent years due to fascinating experimental advances in, e.g., cold-atom physics and photonics.
One common assumption in treating open quantum systems, often referred to as ``factorized initial conditions", is that the system and the environment are initially uncorrelated.  This can be justified for weak system-bath coupling because the system has a negligible impact on
the bath statistics \cite{PollakPRE2008}. On the other hand, for moderate and strong system-bath coupling, which is the case in some realistic situations of great experimental interest (e.g., light-harvesting systems, super-conducting qubits, and atom-cavity systems),
initial system-bath correlations (SBCs) should be accounted for \cite{PechukasPRL,RoyerPRL,Hanggi,TanimuraPRL2010,cheekong}.



 Pioneering studies of SBCs in single-body systems, e.g., a single spin or a single harmonic oscillator in a thermal
 bath, have been fruitful \cite{HakimPRA1985, HaakePRA1985, Grabert1988, SmithPRA1990, GrabertPRE1997, PazPRA1997, LutzPRA2003, BanerjeePRE2003, vanKampen2004, BanPRA2009, UchiyamaPRA, SmirnePRA2010, DajkaPRA, TanPRA2011, MorozovPRA2012, wgwang} (Refs.~\cite{TanimuraPRL2010, ZhangPRA2010} are notable exceptions involving two spins).
The main purpose of this work is to extend the investigation to a system of many identical two-level atoms, where
novel collective phenomena might occur (one example is super-radiance \cite{Dicke1954}).
Specifically, we consider a system of many identical two-level atoms, each atom weakly interacting with a common bath.
Using an exactly solvable model, we show below that effects of
SBCs may dramatically increase with the number of atoms $N$, leading to a previously unknown time scale $t_{c}$
inversely proportional to $N$. For a large system with many atoms, $t_{c}$ becomes very small
and SBCs manifest themselves by inducing rapid oscillations in physical observables.  Three implications of
our findings are in order.
First, the collective dynamics of many two-level atoms can be strongly
affected by SBCs even when each individual two-level atom interacts weakly with the bath.  Second,
this fact may be exploited to amplify and gauge SBCs. Third,
a small $t_{c}$ presents challenges for quantum state protection via dynamical decoupling (DD) techniques.
In general, it is found that SBCs force us to apply more frequent and more efficient DD pulses to freeze the quantum evolution.

In Sec.~II, we present our theoretical findings of SBC effects in a model system consisting
of $N$ two-level atoms in a common bath.   Section III presents computational examples.  We then
discuss in Sec.~IV the implications for the dynamical decoupling technique and how it
may be used to suppress SBC effects.  Section V gives a brief summary.  Readers interested in
the technical details of our derivations should refer to Appendices A-D.

\section{System-bath correlation effects in a system of $N$ two-level atoms in a common bath}
A collection of $N$ identical two-level atoms interacting with a common bosonic bath may be described by
an extended spin-boson Hamiltonian $H_{\text{total}} =  H_S + H_B + H_{\text{int}}$ (setting $\hbar = 1$ throughout) \cite{VorrathChemPhys2004, VorrathPRL2005}, with
\begin{eqnarray}
H_S&=&\omega_0 J_z + \delta J_x + \chi J_z^2; \label{eq1} \\
H_{B}& =& \sum_k \omega_k b_k^\dagger b_k, \\
H_{\text{int}} &= & 2J_z  \sum_k (g_k^* b_k + g_k b_k^\dagger).
\label{totalHamiltonian}
\end{eqnarray}
 Here $J_{x,y,z}$ are the collective spin operators with $J_x^2+J_y^2+J_z^2= \frac{N}{2}(\frac{N}{2}+1)$, $\omega_0$ is the energy bias, $\chi$ the interaction between the atoms, $\delta$ is the tunneling amplitude, and $H_B$ is a collection of boson modes or harmonic oscillators (with zero-point energy dropped).
 Such a total Hamiltonian $H_{\text{total}}$  can model a two-mode BEC interacting via collisions with thermal atoms or phonon excitations \cite{KurizkiPRL2011}.
We stress that, other than the self-interaction term $\chi J_z^2$, $H_{\text{total}}$ is nothing but a straightforward extension of the standard spin-boson model \cite{Weissbook} to a model of many spins interacting with a common bath.

Rather than switching on the system-bath interaction $H_{\text{int}}$ at a particular instant, in most physical situations
the system and the bath have interacted for a long time beforehand. Our starting point is then
a thermal equilibrium state for the system and the bath as a whole at temperature $T$, i.e., $\rho \propto \exp(-\beta H_{\text{total}})$ with $\beta \equiv \frac{1}{k_{b}T}$.
Noticing the energy contribution by $H_{\text{int}}$,
one finds that the associated reduced state of the system (bath) is not really given by $\rho_S^{\text{eq}} \varpropto e^{-\beta H_S}$ ($\rho_B^{\text{eq}}\varpropto e^{-\beta H_B}$) \cite{Weissbook,cheekong}.  Instead, states of the system and of the bath
are correlated for a non-vanishing $H_{\text{int}}$.  Now, if at time $t = 0$, the system is prepared
in a pure state $\ket{\psi}$ via a projective measurement, then the initial state of the system and the bath as a whole is given by
\begin{equation}
\label{initialstate}
\rho(0) = \ket{\psi}\bra{\psi} \otimes \frac{\bra{\psi}e^{-\beta H_{\text{total}}} \ket{\psi}}{Z},
\end{equation}
where $Z$ is a normalization factor. This initial state should be compared with the usual uncorrelated (unphysical) initial state,
\begin{equation}
\label{factorizedinitialstate}
\rho^{\text{dir}}(0) = \ket{\psi}\bra{\psi} \otimes \frac{e^{-\beta H_B}}{\text{Tr}_B(e^{-\beta H_B})}.
\end{equation}
Evidently, for a nonzero $H_{\text{int}}$, we have $\rho(0)\ne \rho^{\text{dir}}(0)$.  That is,
the physical initial state in Eq.~\eqref{initialstate} has correctly accounted for the system-bath interaction during the past.  Consequently
the initial bath state $\rho_{B}(0)=\bra{\psi}e^{-\beta H_{\text{total}}} \ket{\psi}/Z$
depends on $H_{\text{int}}$ as well as the state preparation of the system and is therefore not a canonical equilibrium state for the bath.  Previously,
 how the difference between $\rho(0)$ and $\rho^{\text{dir}}(0)$ impacts on the ensuing dynamics was studied for a damped harmonic oscillator (see, e.g., Ref.~\cite{PazPRA1997}) and a single two-level system undergoing pure dephasing \cite{MorozovPRA2012}. In Appendices C and D, we also consider initial states of the form $\Omega e^{-\beta H_{\text{total}}}\Omega^\dagger/Z$, where $\Omega$ is a unitary operator acting on the system Hilbert space, and we argue that this initial state  for large $N$ leads to similar results as $\rho(0)$ does.

In general, the dynamics of $H_{\text{total}}$ starting from  $\rho(0)$ or $\rho^{\text{dir}}(0)$ does not have analytical solutions. To obtain analytical solutions from which
important insights may be gained, we set the tunneling parameter $\delta$ in $H_S$ to zero, yielding
a pure-dephasing problem for collective spin states.
In the context of a two-mode BEC, such a situation arises if intermode coherent tunneling is made to vanish and if intermode mixing collisions are negligible.
Purely for convenience, we shall assume $\chi=0$, which may be achieved via Feshbach resonance. Up to a unitary transformation, one can obtain equivalent situations if $\omega_0 = 0$ and if inter-mode mixing collisions dominate~\cite{KurizkiPRL2011}.

Working in the basis of $J_z$ eigenstates,  denoted by $|m\rangle$ with $J_z \ket{m} = m\ket{m}$,
we first find the system's reduced density matrix for the uncorrelated initial state $\rho^{\text{dir}}(0)$ (see Appendices for detailed calculations),
\begin{align}
\label{dmuncorrelated-1}
[\rho_S(t)]_{mn} &= [\rho_S(0)]_{mn} e^{-i \omega_0 (m - n) t} e^{-i \Delta(t)(m^2 - n^2)t} \notag \\
&\times e^{-\gamma(t) (m - n)^2 t},
\end{align}
where the decoherence factor is given by
\begin{equation}
\gamma(t)= \sum_k 4|g_k|^2 \frac{(1 - \cos (\omega_k t))}{\omega_k^2 t } \coth \left( \frac{\beta \omega_k}{2} \right),
\end{equation}
and
\begin{equation}
\Delta(t) = \sum_k 4|g_k|^2 \left( \frac{\sin(\omega_k t) - \omega_k t}{\omega_k^2 t}\right).
\end{equation}
The factor $\exp[-i\Delta(t) (m^2-n^2)t]$ arises because the common bosonic bath assists
in generating an indirect atom-atom interaction. Note that at sufficiently short times for which
$\sin(\omega_k t)\approx \omega_k t$, we have $\Delta(t)\approx 0$.

By contrast, for the physical initial state $\rho(0)$ in Eq.~\eqref{initialstate}, we find
\begin{align}
\label{dmcorrelated-1}
[\rho_S(t)]_{mn} &= [\rho_S(0)]_{mn} e^{-i \omega_0 (m - n) t} e^{-i \Delta(t)(m^2 - n^2)t} \notag \\
&\times e^{-\gamma(t) (m - n)^2 t} F^{c}_{mn}(t)
\end{align}
where
\begin{equation}
F^{c}_{mn}(t)=\sum_l p_l e^{-i\left[2l(n - m) \sum_k 4|g_k|^2\sin (\omega_k t)/\omega_k^2\right]},
\end{equation}
with the probability
\begin{equation}
\label{probability}
p_l = \dfrac{|\langle l | \psi\rangle |^2 e^{-\beta \omega_0 l} e^{\beta l^2 \mathcal{C}}}{\sum_l \left( |\langle l | \psi\rangle |^2 e^{-\beta \omega_0 l} e^{\beta l^2 \mathcal{C}}\right)},
\end{equation}
and
\begin{equation}
\mathcal{C} = \sum_k  4|g_k|^2/\omega_k.
\end{equation}
Comparing  Eq.~\eqref{dmuncorrelated-1} with Eq.~\eqref{dmcorrelated-1}, it is seen that
effects of the SBC on the dynamics are entirely captured by the factor $F^c_{mn}(t)$.
To understand this, we first examine the non-canonical bath state $\rho_B(0)$, i.e., the reduced bath state at time zero.
Up to a normalization factor, we obtain
\begin{equation}
\rho_B(0) \propto \sum_l e^{-\beta \omega_0 l} |\ip{l}{\psi}|^2 e^{-\beta H_B^{(l)}},
\end{equation}
 where
 \begin{equation}
 H_B^{(l)} = H_B + 2l \sum_k (g_k^* b_k + g_k b_k^\dagger).
 \end{equation}
 Two observations can be made here. First, the bath is prepared in the state $e^{-\beta H_B^{(l)}}$ with a probability
related to the system projection amplitude $\ip{l}{\psi}$.  Second, $H_B^{(l)}$ can be interpreted as a collection of harmonic oscillators, each of which is under a  `force' proportional to $2l$. Due to this force exerted by the system, the actual equilibrium position of the
bath oscillators will be displaced.  To make this clearer, we define displaced harmonic oscillator modes
\begin{equation}
B_{k,l} \equiv b_k + 2lg_k/\omega_k,
\end{equation}
from which we have
\begin{equation}
\rho_B(0) \varpropto \sum_l e^{-\beta \omega_0 l} |\ip{l}{\psi}|^2 e^{\beta l^2 \mathcal{C}} e^{-\beta  \sum_k \omega_k B_{k,l}^\dagger B_{k,l}}.
 \end{equation}
 The initial bath state is seen to be a mixture of different components: each component
is a canonical equilibrium state $\varpropto e^{-\beta  \sum_k \omega_k B_{k,l}^\dagger B_{k,l}}$ for a collection of harmonic oscillators displaced by $2lg_k/\omega_k$, with the probability $p_l$ defined in Eq.~\eqref{probability}.


An interesting physical picture then arises. At time zero a projection of the system on state $|\psi\rangle$
breaks the equilibrium state of the system and the bath as a whole and the bath is instead prepared
in a mixture of many components $\sim e^{-\beta  \sum_k  \omega_k B_{k,l}^\dagger B_{k,l}}$,
with each component assuming a certain $J_z$ eigenstate for the system.
For the $|m\rangle$ component of $|\psi\rangle$, the component $e^{-\beta  \sum_k  \omega_k B_{k,l}^\dagger B_{k,l}}$ of $\rho_B(0)$
finds its equilibrium condition no longer satisfied for $l\ne m$.  As a result the collection of displaced harmonic modes start to oscillate
around their new equilibrium positions defined by $2m g_k/\omega_k$.  For another $|n\rangle$ component of $|\psi\rangle$,
the same mechanism works but with a different degree due to new equilibrium positions at $2n g_k/\omega_k$.  It is such type of bath motion
that yields the correction factor $F^c_{mn}(t)$ in Eq.~(\ref{dmcorrelated-1}).

Because $e^{\beta l^2\mathcal{C}}$ changes rapidly with $l\in [-N/2, N/2]$, for large $N$
only the terms with probabilities $p_{\pm \frac{N}{2}}$ may contribute to $F^{c}_{mn}(t)$. Further, for $N \beta \omega_0 \gg 1$, we have
$e^{N\beta \omega_0/2} \gg e^{-N\beta \omega_0/2}$ and hence $p_{-\frac{N}{2}} \gg p_{\frac{N}{2}}$ for a generic state $|\psi\rangle$.  This
yields $p_{-\frac{N}{2}}\approx 1$ in our many-body system, thus reducing $F^{c}_{mn}(t)$ to
\begin{equation}
F^c_{mn}(t)\approx e^{i\left[ N (n - m) \sum_k 4|g_k|^2\sin (\omega_k t)/\omega_k^2\right]}.
\end{equation}
In particular, at sufficiently short times,  $\sin (\omega_k t) \approx \omega_k t$
and then $ F^c_{mn}(t)\approx e^{i \left[N (n - m) t \sum_k 4|g_k|^2 /\omega_k \right]}$.  That is, $F^c_{mn}$ represents
a phase factor building up with time at a rate of  $ N (n - m)  \sum_k 4|g_k|^2 /\omega_k$. Note that during the same time window,
$\Delta(t)$ originating from bath-assisted atom-atom interaction
still stays close to zero due to $\sin (\omega_k t) \approx \omega_k t$.  For physical observables not diagonal
in the $|m\rangle$ representation, the time dependence of $F^c_{mn}$
leads to a characteristic time scale (after setting $|m-n|=1$):
\begin{equation}
t_c= \left(N\sum_k 4|g_k|^2 /\omega_k\right)^{-1}.
\end{equation}
Remarkably, $t_c$ is inversely proportional to the number of two-level atoms.  As an example we consider an Ohmic spectral density for the bath, with
\begin{equation}
\sum_k 4|g_k|^2 C(\omega_k)\rightarrow \int^{\infty}_{0} d\omega J(\omega)C(\omega)
\end{equation} and
\begin{equation}
J(\omega)=G\omega e^{-\omega/\omega_c}.
\end{equation}
 Then
one finds $t_c= (NG\omega_c)^{-1}$, a time scale determined by parameters from both the system ($N$) and the bath ($G$ and $\omega_c$).
On the other hand, at later times, the time dependence of $F^{c}_{mn}(t)$ weakens as the oscillations of the bath oscillators around their new
equilibrium positions start to dephase. For the Ohmic spectrum, $F^{c}_{mn}(t)$ at long times is found to approach
$e^{i\left[ \frac{\pi}{2} NG(n-m)\right]}$. Analogous results are found for other spectral density functions $J(\omega)$ with an exponential cutoff.

\section{Numerical examples of amplified system-bath correlation effects}

\begin{figure}[b]
   \includegraphics[scale = 1.08]{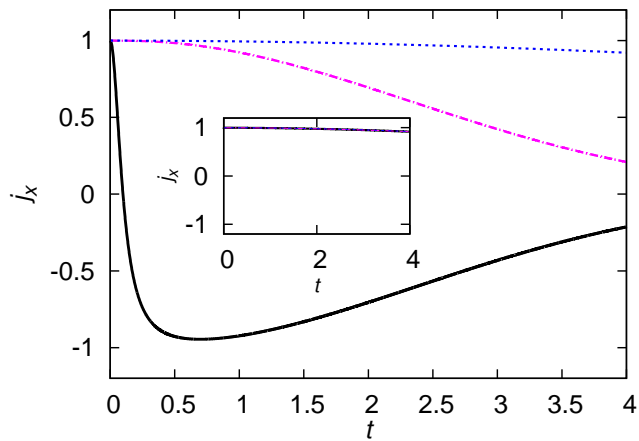}
   \centering
  	\caption{(color online) $j_x$ vs $t$ without a bath (upper dotted, blue), with an Ohmic bath but using a factorized initial state (dot-dashed, magenta), and with an Ohmic bath and including SBC effects (solid, black). $N = 2000$, $\omega_0 = 0.1$, $ G = 0.001$, $\beta = 1000$, and $\omega_c = 10$ \cite{footnoteparameters}. Inset shows the parallel results if $N=1$, where three lines become almost indistinguishable.}
  	\label{graphJ1000andJ0p5}
\end{figure}

\begin{figure}[t]
   \includegraphics[scale = 1.15]{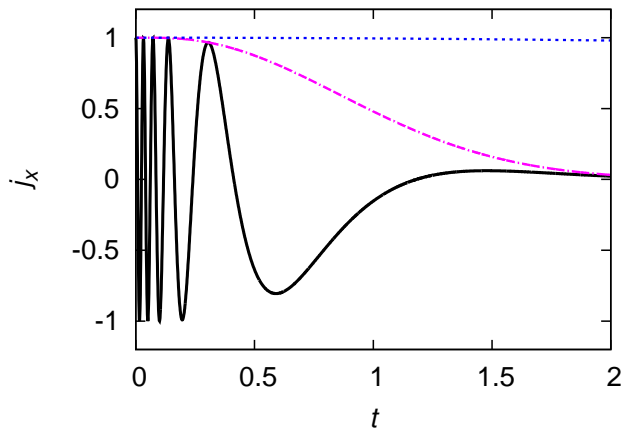}
   \centering
  	\caption{(color online) Same as in Fig.~\ref{graphJ1000andJ0p5}, but now with $N = 20000$.  The rapid oscillations in $j_x$ clearly demonstrate the
  amplification of SBC effects achieved by an increase in the number of particles.}
  	\label{graphJ10000}
\end{figure}

In this section we turn to a concrete example for which we investigate the dynamics of a scaled observable
$j_x \equiv 2\langle J_x \rangle / N$.  At time zero, the system is
projected onto the initial spin state $\ket{\psi} = e^{-i\frac{\pi}{2} J_y} \ket{N/2}$.
Using the uncorrelated initial state in Eq.~(\ref{factorizedinitialstate}), we obtain
\begin{equation}
j_x = e^{-\gamma(t)t}\text{Re}\left[ e^{i[\omega_0 + \Delta(t)]t} X(t) \right],
\end{equation} with
\begin{eqnarray}
X(t) &=&  \frac{2}{N}\sum_m \frac{1}{2^N} e^{2imt\Delta(t)} \nonumber \\
 &&\times\ \sqrt{\binom{N}{\frac{N}{2}+m} \binom{N}{\frac{N}{2}+m+1} \left(\frac{N}{2} - m\right) \left(\frac{N}{2} + m + 1\right)}.
\end{eqnarray}
Details can be found in Appendix D.  By contrast, the true dynamics with the physical initial state in Eq.~(\ref{initialstate}) gives
\begin{equation}
j_x = e^{-\gamma(t)t}\text{Re}\left[ e^{i[\omega_0 + \Delta(t)]t} X(t) F^{c}_{x}(t) \right],
\end{equation} with $F^c_x(t)\approx e^{iNG\omega_c t}$ at short times
for an Ohmic spectrum defined above. In Fig.~\ref{graphJ1000andJ0p5}, we plot the time dependence of $j_x$
for $N = 2000$, in the absence or presence of a bath with $ G \ll 1$  (so that
each individual two-level atom interacts with the bath very weakly \cite{VorrathChemPhys2004}).
For the shown time period in Fig.~\ref{graphJ1000andJ0p5},  $j_x$ in the absence of the bath
hardly changes due to a finite $\omega_0$. In the presence of the bath but without including SBCs, $j_x$ stays positive but decreases at a faster rate due to the bath-assisted atom-atom interaction.
With SBCs accounted for, completely different qualitative behavior of $j_x$ is observed:
it rapidly changes from $1$ to $-1$ and then gradually returns to a value close to zero. Note that,
as analyzed in theory, such rapid change in $j_x$ occurs
before atom-atom indirect interaction (as captured by $\Delta (t)$) takes effect.
The inset also shows the parallel results if $N=1$, with all other parameters unchanged.  As expected from $G\ll 1$,
in that case all lines are almost on top of each other and hence no SBC effect can be seen.
Because the decoherence function $\gamma(t)$ is independent of $N$,
the results in the inset also hint that decoherence for the $N=2000$ case is insignificant for the shown time scale.
Therefore, SBCs impact on the dynamics long before the onset of decoherence.

To emphasize the role of $N$, in Fig.~\ref{graphJ10000} we plot parallel results for $N = 20000$.
There SBCs induce more drastic oscillations in $j_x$, followed by slower oscillations
as the time dependence of $F^{c}_{mn}(t)$ weakens. Clearly then, at short times
an increasing number of atoms enhances the oscillation frequency in physical observables,
thus amplifying the SBC effects.


\section{Suppression of system-bath correlation effects by dynamical decoupling}
In this section we discuss the implications of this work for DD, which is one main approach to quantum state protection \cite{LloydDD, UhrigDD,liuUDD}.
 DD in single-spin systems has found enormous applications. It is therefore highly desirable to extend DD to systems describable by collective (large) spins $J_{x,y,z}$ \cite{adampra2012,largespin}.  We return to our pure-dephasing model introduced in Sec.~II.
Because the details of a bath spectrum is unknown in general, the precise form of
the correction factor $F^c_{mn}(t)$ induced by SBCs or of the $\Delta(t)$-related phase factor due to bath-assisted atom-atom interaction is unavailable
in general. It is tempting to wait for $F^{c}_{mn}(t)$ to saturate
such that its time dependence is out of the picture. However, as indicated by our theory and shown in Fig.~\ref{graphJ10000},
before reaching that regime the bath assisted atom-atom interaction would have already changed the state.
So in order to protect or store a given many-body state, the time dependence of $F^c_{mn}(t)$ and $\Delta(t)$
must be suppressed.
In general the task of DD becomes three-fold: to suppress the decoherence factor $\gamma(t)$, the $\Delta(t)$-related phase factor, and the correction
factor $F^c_{mn}(t)$.  For large $N$, it is found that
DD control pulses need to first compete with the correlated-induced time scale $t_c\varpropto 1/N$.

Consider now $N_d$
instantaneous $\pi$-pulses of $J_x$ applied to our system at times $t_l$, with $1\leq l\leq N_d$.  Upon application of
one such pulse, one has, in the frame of the applied pulses, $J_z\rightarrow -J_z$.  It is hence convenient to introduce the so-called switching function
$f(t)$, with
\begin{equation}
f(t) = \sum_{l = 1}^{N_d + 1} (-1)^{l + 1} \theta(t - t_{l - 1})\theta(t_l - t),
\end{equation}
where $\theta(t)$ is the Heaviside function.  With the assistance of $f(t)$,
one can still find the time dependence of $[\rho_{S}(t)]_{mn}$ or of an observable such as $j_x$
analytically, using the same technique as used in previous cases without DD pulses.
Take $[\rho_{S}(t)]_{mn}$ under DD as an example. The general form
of Eq.~(\ref{dmcorrelated-1}) still holds, but now with $\omega_0(t)$ changed to
$\tilde{\omega}_0(t)$, $\Delta(t)$ changed to $\tilde{\Delta}(t)$, $\gamma(t)$ changed to
$\tilde{\gamma}(t)$, and $F_{mn}^{c}(t)$ changed to $\tilde{F}^{c}_{mn}(t)$.  Specifically,
we find
\begin{eqnarray}
\tilde{\omega}_0 & = & \frac{\omega_0}{t} \int_0^t f(t') dt', \nonumber \\
\tilde{\Delta}(t)&  = & \sum_k \frac{4|g_k|^2}{t} \int_0^t dt_1\int_0^{t_1}dt_2 f(t_1)f(t_2) \sin [\omega_k (t_2 - t_1)], \nonumber
 \\
\tilde{\gamma}(t) &= & \sum_k 4|g_k|^2 \frac{|f(\omega_k,t)|^2}{\omega_k^2 t } \coth \left( \frac{\beta \omega_k}{2} \right),
\end{eqnarray}
and
\begin{equation}
\tilde{F}^{c}_{mn}(t)  \approx   e^{-i\left[ N (n - m) \sum_k 4|g_k|^2\text{Im}[f(\omega_k,t)]/\omega_k^2\right]}.
\label{feq}
\end{equation}
Here
\begin{equation}
f(\omega_k,t)= 1 + (-1)^{N_d + 1} e^{i\omega_k t} + 2\sum_{l = 1}^{N_d} (-1)^l e^{i\omega_k t_l}
\end{equation}
is determined by $f(t)$ via the following relation
\begin{equation}
f(\omega_k, t) = -i \omega_k \int_0^t dt' e^{i\omega_k t'} f(t').
\end{equation}
With these explicit results, effects of DD pulses on the dynamics with SBC effects can be examined in detail.
Of particular interest is $\tilde{F}^{c}_{mn}(t)$ because it is indicative of the impact of an applied DD sequence
on SBC effects at early times.

\begin{figure}[t]
   \includegraphics[scale = 1.15]{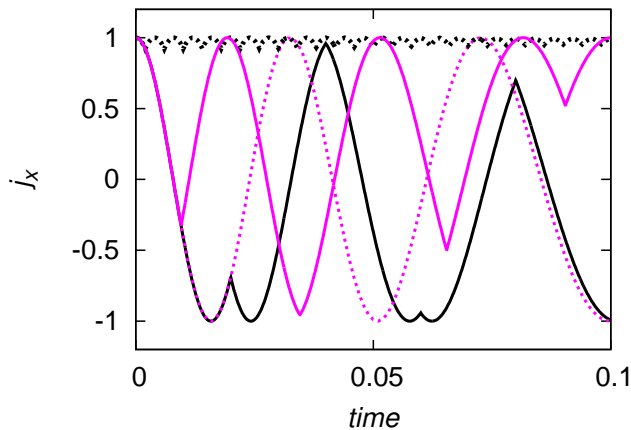}
   \centering
  	\caption{(color online) $j_x$ vs time with SBC effects and under bang-bang control pulses, for $\tau = 0.02$ (solid, black) or $\tau = 0.002$ (dotted, black). The parallel results without any control pulses (dotted, magenta) and with four UDD pulses (solid, magenta) are also shown.
  Parameters are the same as in Fig.~\ref{graphJ10000}, with the final time chosen as $t=0.1$.}
  	\label{graphwithpulses}
\end{figure}

In Fig.~~\ref{graphwithpulses} we illustrate the effect of applying DD pulses for $N=20000$
for a final time $t = 0.1$.
We first investigate the usefulness of equidistance control pulses (bang-bang control) with a pulse interval $\tau$ \cite{LloydDD}.
Note first that if we neglect SBCs, then there would be no need to apply any DD pulses
because the evolution of $j_x$ is negligible for the considered period.  With SBCs accounted for,
an application of $N_d = 4$ pulses with $\tau=0.02$ is found to be insufficient for state protection.
Indeed, for the system parameters used we find
$t_c= (GN\omega_c)^{-1}=0.005$, which is far smaller than $\tau=0.02$.  To compete with $t_c$, we then use $\tau = 0.002<t_c$, for which
the evolution is successfully frozen (see Fig.~\ref{graphwithpulses}).

To avoid the need of a high pulse repetition rate,
we next consider the celebrated, more effective, Uhrig's dynamical decoupling (UDD) sequence \cite{UhrigDD,liuUDD} with unequal pulse intervals, where the pulse timings are given by $t_j = t \sin^2 \left( \frac{j\pi}{2N_p + 2} \right)$.
Dramatically, via only $N_d=4$ UDD pulses, the $j_x$ value at the final time $t=0.1$ is already recovered to its initial value.  A careful
look into the above expression for $\tilde{F}^{c}_{mn}(t)$ explains why this is so.
The original motivation of a UDD sequence is to suppress the decoherence function $\gamma(t)$ by minimizing $|f(\omega_k,t)|$
to its $N_d$-th order in time.  So by construction, the time evolution of $\tilde{F}^{c}_{mn}(t)$ is also optimally suppressed
by UDD because a minimized $|f(\omega_k,t)|$ automatically yields a minimized $\text{Im}[f(\omega_k,t)]$ that enters into  $\tilde{F}^{c}_{mn}(t)$ in Eq.~(\ref{feq}).
That is, here
the correction factor $\tilde{F}_{mn}^{c}(t)$ is already optimally suppressed for an unknown spectrum with an exponential cutoff.
Additional computational studies indicate that
the $\tilde{\Delta}(t)$ term may be also well suppressed by DD pulses, though this is not of interest here as our main concern is to
suppress SBC before bath-assisted atom-atom interaction has any considerable effect on the dynamics.
Our conclusions are as follows.
In locking a many-body state in our model here,  the previously unknown time scale $t_c$ induced by SBC
calls for the use of DD control pulses long before decoherence and bath-assisted atom-atom interaction becomes important,
with a UDD sequence found to be an optimized choice (assuming the detailed form of the bath spectrum is not available).

\section{summary}
In summary, because SBC effects are amplified by the number of particles
in a many-body system, they can be important
even when each individual particle interacts with the bath weakly. Interestingly,
though SBC effects originate from system-bath interaction in the long past, they may still
be suppressed by dynamical decoupling so that a prepared many-body state is well protected.  Nevertheless,
reaching this goal calls for more effective control pulses applied within a shorter time scale.
Our results should also be of interest to other subtopics in open many-body systems by
considering SBC effects neglected before.

\section{Acknowledgment}
We would like to thank Derek Ho for insightful discussions.

\appendix

\section{Exact unitary evolution operator}

The dynamics for our model described by Eqs.~(\ref{eq1})-(\ref{totalHamiltonian}) can be exactly solved if  $\delta = 0$.
Purely for convenience we also assume $\chi = 0$.  We first transform to the interaction picture.  In this picture,
the Hamiltonian becomes
\begin{align}
H_I(t) &= e^{i(H_S + H_B)t} H_{\text{int}} e^{-i(H_S + H_B)t}, \notag \\
&= 2J_z \sum_k (g_k^* b_k e^{-i\omega_k t} + g_k b_k^\dagger e^{i\omega_k t} ).
\end{align} Similar to the treatment in Ref.~\cite{KurizkiPRL2011}, we next find the time evolution operator $U_I(t)$ corresponding to $H_I(t)$. The Magnus expansion \cite{Magnus1954} tells us that
\begin{equation}
U_I(t) = \exp \left[ \sum_{i = 1}^{\infty} A_i(t) \right],
\end{equation}
where
\begin{eqnarray}
A_1(t) = -i \int_0^t dt_1 H_I(t_1), \\
A_2(t) = -\frac{1}{2} \int_0^t dt_1 \int_0^{t_1} dt_2 [H_I(t_1),H_I(t_2)].
\end{eqnarray}
It is straightforward to find
\begin{align}
A_1 &= -i \int_0^t dt_1 H_I(t_1), \notag \\
&= J_z \sum_k (b_k^\dagger \alpha_k(t) - b_k \alpha_k^*(t)),
\end{align}
where
\begin{equation}
\alpha_k(t) = \frac{2g_k (1 - e^{i\omega_k t})}{\omega_k}.
\end{equation}
Further using
\begin{equation}
[H_I(t_1),H_I(t_2)] =  -8iJ_z^2 \sum_k |g_k|^2 \sin[\omega_k (t_1 - t_2)],
\end{equation}
we obtain
\begin{align}
A_2 &= \frac{1}{2} \int_0^t dt_1 \int_0^{t_1} dt_2 8i J_z^2 \sum_k |g_k|^2 \sin [\omega_k(t_1 - t_2)], \notag \\
&= -iJ_z^2t\Delta(t),
\end{align}
where
\begin{equation}
\Delta(t) \equiv \frac{1}{t} \sum_k 4|g_k|^2 \frac{[\sin(\omega_k t) - \omega_k t]}{\omega_k^2}.
\end{equation}
Since this is a c-number, the higher order terms in the Magnus expansion are all zero.  The exact unitary time evolution operator is hence found, i.e.,
\begin{equation}
U(t) = e^{-i\omega_0 J_z t} e^{-iH_B t} U_I(t),
\end{equation}
where
\begin{equation}
U_I(t) = \exp \lbrace J_z \sum_k [b_k^\dagger \alpha_k(t) - b_k \alpha_k^*(t)] - iJ_z^2 t \Delta(t) \rbrace.
\end{equation}

For later calculations, let us first consider the reduced density operator of the system
\begin{equation}
\rho_S(t) = \text{Tr}_B [U(t) \rho(0) U^\dagger (t) ],
\end{equation}
where $\rho(0)$ is the density operator of the system and the bath as a whole.  We find it useful
to write the reduced density operator of the system in terms of the standard $J_z$ basis as
\begin{equation}
[\rho_S(t)]_{mn} = \text{Tr}_{S,B} [U(t) \rho(0) U^\dagger (t) P_{nm}].
\end{equation}
Here $P_{nm} \equiv \ket{n}\bra{m}$, $\ket{n}$ being the eigenstate of $J_z$ with eigenvalue $n$. Introducing the Heisenberg picture operator $P_{nm}(t)$ as $U^\dagger(t) P_{nm} U(t)$, we have
\begin{equation}
[\rho_S(t)]_{mn} = \text{Tr}_{S,B} [P_{nm}(t) \rho(0)].
\end{equation}
Calculations of the explicit form of $P_{nm}(t)$ are straightforward because the time evolution operator $U(t)$ is already found.  In particular, we obtain
\begin{equation}
P_{nm}(t) = e^{-i\omega_0 t(m-n)} e^{-i\Delta(t) t (m^2 - n^2)}  e^{-R_{nm}(t)} P_{nm},
\end{equation}
where
\begin{equation}
R_{nm}(t) = (n - m) \sum_k [b_k^\dagger \alpha_k(t) - b_k \alpha_k^*(t)].
\end{equation}
It follows that
\begin{eqnarray}
\label{putininitialstatehere}
[\rho_S(t)]_{mn}=  e^{-i\omega_0 t (m - n)} e^{-i \Delta(t)t(m^2 - n^2)}  \text{Tr}_{S,B} [e^{-R_{nm}(t)} P_{nm} \rho(0)].
\end{eqnarray}
This is a general result because it applies to an arbitrary initial density $\rho(0)$ (the unphysical
state $\rho^{\text{dir}}(0)$ or the physical state with SBC accounted for).

\section{Dynamics with uncorrelated initial states}
Here we consider unphysical decorrelated initial states, i.e.,
\begin{equation}
\rho^\text{dir}(0) = \rho_S(0) \otimes \rho_B,
\end{equation}
where $\rho_B = \frac{e^{-\beta H_B}}{Z_B}$ with $Z_B = \text{Tr}_B [e^{-\beta H_B}]$. Then,
\begin{eqnarray}
[\rho_S(t)]_{mn}  =  [\rho_S(0)]_{mn} e^{-i\omega_0 (m - n) t} e^{-i\Delta(t)(m^2 - n^2)t} \text{Tr}_B [e^{-R_{nm}(t)} \rho_B].
\end{eqnarray}
We now simplify $\text{Tr}_B[e^{-R_{nm}(t)} \rho_B] = \langle e^{-R_{nm}(t)}\rangle$, where the average is taken with respect to the thermal bath state at equilibirum. Although this is a standard result, for self-completeness, we explain the steps in detail. Since the modes are independent of each other, we can write
\begin{equation}
\langle e^{-R_{nm}(t)}\rangle = \prod_k \langle e^{-(n - m) [b_k^\dagger \alpha_k(t) - b_k \alpha_k^*(t)]} \rangle.
\end{equation}
For an operator $A$ which is a linear combination of creation and annihilation operators, we have $\langle e^A \rangle = e^{\langle A^2 \rangle/2}$. Using this identity, we have
\begin{align}
\langle e^{-R_{nm}(t)}\rangle &= \prod_k \exp \left[ -\frac{1}{2} (n - m)^2 |\alpha_k(t)|^2 \langle b_k^\dagger b_k + b_k b_k^\dagger \rangle \right] \notag \\
&= \prod_k \exp \left[ -\frac{1}{2} (n - m)^2 |\alpha_k(t)|^2 \langle 2n_k + 1 \rangle \right].
\end{align}
Further using the definition of $\alpha_k(t)$ and the Bose-Einstein distribution, we find that
\begin{eqnarray}
 \text{Tr}_B[e^{-R_{nm}(t)} \rho_B] \nonumber = \exp \left[ -\sum_k (m - n)^2 4 |g_k|^2 \frac{[1 - \cos (\omega_k t)]}{\omega_k^2} \coth \left(\frac{\beta \omega_k}{2} \right) \right].
\end{eqnarray}
This then yields
\begin{equation}
[\rho_S(t)]_{mn} = [\rho_S(0)]_{mn} e^{-i \omega_0 (m - n) t} e^{-i \Delta(t)(m^2 - n^2)t} e^{-\gamma(t) (m - n)^2 t},
\end{equation}
with
\begin{equation}
\gamma(t) = \frac{1}{t} \sum_k 4|g_k|^2 \frac{(1 - \cos (\omega_k t))}{\omega_k^2} \coth \left( \frac{\beta \omega_k}{2} \right).
\end{equation}
The factor $e^{-\gamma(t) (m - n)^2 t}$ describes decoherence and the factor $e^{-i \Delta(t)(m^2 - n^2)t}$ describes
the indirect atom-atom interaction induced by the common bath.

\section{Dynamics with system-bath-correlated initial states}
Here we consider (physical) correlated initial states of the form
\begin{equation}
\label{correlatedinitialstate}
\rho(0) = \frac{1}{Z} \sum_r \Omega_r e^{-\beta H_{\text{total}}} \Omega_r^\dagger,
\end{equation}
with
\begin{equation}
Z = \text{Tr}_{S,B} \left[\sum_r \Omega_r e^{-\beta H_{\text{total}}} \Omega_r^\dagger \right].
\end{equation}
The $\Omega_r$ operators above are assumed to be acting on the system only and their
explicit forms will be specified later. To solve for the dynamics starting from such an initial state, we may use a polaron transformation technique, generalizing the results of Ref.~\cite{MorozovPRA2012} from a single spin to many spins, or we may use displaced harmonic oscillator modes. The latter method is used below
 since it is physically more transparent.

We first simplify $Z$. By introducing a completeness relation, we find that
\begin{align}
Z &= \sum_r \sum_l \text{Tr}_{S,B} [ \Omega_r e^{-\beta H_{\text{total}}} \ket{l}\bra{l} \Omega_r^\dagger ], \notag \\
&= \sum_r \sum_l e^{-\beta \omega_0 l} \opav{l}{\Omega_r^\dagger \Omega_r}{l} \text{Tr}_B [e^{-\beta H_B^{(l)}}],
\end{align}
where we have defined
\begin{equation}
H_B^{(l)} = H_B + 2l \sum_k (g_k^* b_k + g_k b_k^\dagger).
\end{equation}
Using the displaced harmonic oscillator modes,
\begin{align}
B_{k,l} = b_k + \frac{2lg_k}{\omega_k}, \\
B_{k,l}^\dagger = b_k^\dagger + \frac{2lg_k^*}{\omega_k}.
\end{align}
we find
\begin{equation}
Z = \sum_r \sum_l e^{-\beta \omega_0 l} \opav{l}{\Omega_r^\dagger \Omega_r}{l} e^{\beta l^2 \mathcal{C}} Z_B,
\end{equation}
where $\mathcal{C} = \sum_k \frac{4|g_k|^2}{\omega_k}$, and in this case $Z_B = \text{Tr}_B [e^{-\beta \sum_k \omega_k B_{k,l}^\dagger B_{k,l}} ]$.

We now substitute Eq.~\eqref{correlatedinitialstate} in Eq.~\eqref{putininitialstatehere} and again introduce a completeness relation. To proceeed, $\text{Tr}_B [ e^{-R_{nm} (t)} e^{-\beta H_B^{(l)}} ]$ needs to be simplified. As before, this can be done using displaced oscillator modes. It is straightforward to show that
\begin{equation}
R_{nm}(t) = (n - m) \sum_k [\alpha_k (t) B_{k,l}^\dagger - \alpha_k^* (t) B_{k,l} ] + i\Phi_{nm}^{(l)} (t),
\end{equation}
where
\begin{eqnarray}
\Phi_{nm}^{(l)} & =& 2(n - m)l \Phi(t), \\
\Phi(t) & = & \sum_k \frac{4|g_k|^2}{\omega_k^2} \sin (\omega_k t).
\end{eqnarray}
We then find that
\begin{equation}
\text{Tr}_B [e^{-R_{nm}(t)} e^{-\beta H_B^{(l)}}] = e^{-i\Phi_{nm}^{(l)}(t)} e^{\beta l^2 \mathcal{C}} Z_B e^{-\gamma(t) (m - n)^2 t},
\end{equation} which leads to
\begin{eqnarray}
[\rho_S(t)]_{mn}& =&  e^{-i\omega_0 (m - n)t} e^{-i\Delta(t)(m^2 - n^2)t}  e^{-\gamma(t)(m - n)^2 t} \nonumber \\
&&\times\ \dfrac{\sum_r \sum_l \left( \opav{l}{\Omega_r^\dagger P_{nm} \Omega_r}{l} e^{-i\Phi_{nm}^{(l)}(t)} e^{-\beta \omega_0 l} e^{\beta l^2 \mathcal{C}}\right)}{\sum_r \sum_l \left( \opav{l}{\Omega_r^\dagger \Omega_r}{l} e^{-\beta \omega_0 l}  e^{\beta l^2 \mathcal{C}}\right)}.
\end{eqnarray}
Noting that
\begin{equation}
[\rho_S(0)]_{mn} = \dfrac{\sum_r \sum_l \left( \opav{l}{\Omega_r^\dagger P_{nm} \Omega_r}{l}  e^{-\beta \omega_0 l} e^{\beta l^2 \mathcal{C}}\right)}{\sum_r \sum_l \left( \opav{l}{\Omega_r^\dagger \Omega_r}{l} e^{-\beta \omega_0 l}  e^{\beta l^2 \mathcal{C}}\right)},
\end{equation}
one finally arrives at
\begin{align}
[\rho_S(t)]_{mn} &= [\rho_S(0)]_{mn} e^{-i\omega_0 (m - n)t} e^{-i \Delta(t)(m^2 - n^2)t}  e^{-\gamma(t)(m - n)^2 t} \notag \\
&\times \dfrac{\sum_r \sum_l \left( \opav{l}{\Omega_r^\dagger P_{nm} \Omega_r}{l} e^{-i\Phi_{nm}^{(l)}(t)} e^{-\beta \omega_0 l} e^{\beta l^2 \mathcal{C}}\right)}{\sum_r \sum_l \left( \opav{l}{\Omega_r^\dagger P_{nm} \Omega_r}{l} e^{-\beta \omega_0 l} e^{\beta l^2 \mathcal{C}}\right)}.
\end{align}

\subsection{State preparation via projective measurement}

In this case,
\begin{equation}
\label{preparationviameasurement}
\rho(0) = \frac{1}{Z} P_{\psi} e^{-\beta H_{\text{total}}} P_{\psi}
\end{equation}
with $P_{\psi} = \ket{\psi}\bra{\psi}$.  Using our general results above we obtain
\begin{eqnarray}
\label{dmcorrelated}
[\rho_S(t)]_{mn}& = &[\rho_S(0)]_{mn} e^{-i\omega_0 (m - n)t} e^{-i \Delta(t)(m^2 - n^2)t}  e^{-\gamma(t)(m - n)^2 t} \nonumber \\
&& \times \ \dfrac{\sum_l \left( |\langle l | \psi\rangle |^2 e^{-i\Phi_{nm}^{(l)}(t)} e^{-\beta \omega_0 l}  e^{\beta l^2 \mathcal{C}}\right)}{\sum_l \left( |\langle l | \psi\rangle |^2 e^{-\beta \omega_0 l}  e^{\beta l^2 \mathcal{C}}\right)}.
\end{eqnarray}
This can be rewritten as
\begin{equation}
[\rho_S(t)]_{mn} = [\rho_S(0)]_{mn} e^{-i \omega_0 (m - n) t} e^{-i \Delta(t)(m^2 - n^2)t}
e^{-\gamma(t) (m - n)^2 t} F_{mn}^c(t),
\end{equation}
where $F_{mn}^c(t)$ is already given in the main text.

\subsection{State preparation via unitary operations}
Alternatively, instead of performing a projective measurement on the system, we may first cool the system and the bath to a desired low temperature.
We then perform a unitary operation on the system to approximately arrive at some initial state.  In general, initial state of the system and the bath prepared in this manner can be written as
\begin{equation}
\rho(0) = \frac{1}{Z} \Omega e^{-\beta H_{\text{total}}} \Omega^\dagger,
\end{equation}
with $\Omega$ being a unitary preparation operator acting on the system Hilbert space. We then have
\begin{eqnarray}
\label{dmcorrelatedunitary}
[\rho_S(t)]_{mn} & = & [\rho_S(0)]_{mn} e^{-i\omega_0 (m - n)t} e^{-i \Delta(t)(m^2 - n^2)t}  e^{-\gamma(t)(m - n)^2 t} \nonumber \\
&& \times\ \dfrac{\sum_l \left( \opav{l}{\Omega^\dagger}{n} \opav{m}{\Omega}{l} e^{-i\Phi_{nm}^{(l)}(t)} e^{-\beta \omega_0 l}  e^{\beta l^2 \mathcal{C}}\right)}{\sum_l \left( \opav{l}{\Omega^\dagger}{n} \opav{m}{\Omega}{l} e^{-\beta \omega_0 l}  e^{\beta l^2 \mathcal{C}}\right)}.
\end{eqnarray}
This can be rewritten as
\begin{equation}
[\rho_S(t)]_{mn} = [\rho_S(0)]_{mn} e^{-i\omega_0 (m - n)t} e^{-i \Delta(t)(m^2 - n^2)t}  e^{-\gamma(t)(m - n)^2 t} \left[ \sum_l \mu^{(l)}_{nm} e^{-i \Phi_{nm}^{(l)}(t)} \right],
\end{equation}
with
\begin{equation}
\mu^{(l)}_{nm} = \dfrac{\opav{l}{\Omega^\dagger}{n} \opav{m}{\Omega}{l} e^{-\beta \omega_0 l}  e^{\beta l^2 \mathcal{C}}}{\sum_l \left( \opav{l}{\Omega^\dagger}{n} \opav{m}{\Omega}{l} e^{-\beta \omega_0 l}  e^{\beta l^2 \mathcal{C}}\right)}.
\end{equation}
Note that $\mu^{(l)}_{nm}$ need not be real here (nevertheless, by its definition only, we still have $\sum_l \mu^{(l)}_{nm} = 1$).

\section{Calculating observables}
To evaluate physical observables using the system's reduced density operator, let us first evaluate the functions $\gamma(t)$, $\Phi(t)$, $\mathcal{C}$ and $\Delta(t)$. As usual, we take the continuum limit of the bath modes, whereby the summations over the bath modes are replaced by integrals via the rule
\begin{equation}
\sum_k 4 |g_k|^2 C(\omega_k) = \int_0^\infty d\omega J(\omega) C(\omega),
\end{equation}
with $J(\omega)$ being the spectral density. For this work, we choose Ohmic spectral density, that is,
\begin{equation}
J(\omega) = G \omega e^{-\omega/\omega_c}.
\end{equation}
We can then evaluate the integrals as \cite{MorozovPRA2012},
\begin{align}
\mathcal{C} = G \omega_c, \\
\Phi(t) = G \tan^{-1} (\omega_c t),
\end{align}
and
\begin{align}
\gamma(t) &= \gamma_{\text{vac}}(t) + \gamma_{\text{th}}(t), \\
\gamma_{\text{vac}}(t) &= \frac{G}{2t} \text{ln} (1 + \omega_c^2 t^2), \\
\gamma_{\text{th}}(t) &= \frac{2G}{t} [ \text{ln} \Gamma(1 + \frac{1}{\beta \omega_c}) - \frac{1}{2} \text{ln} |\Gamma(1 + \frac{1}{\beta \omega_c} + \frac{it}{\beta})|^2], \\
\Delta(t) &= \frac{1}{t} (\Phi(t) - \mathcal{C}t).
\end{align}
Expressions for sub-Ohmic and super-Ohmic spectral densities can also be found in Ref.~\cite{MorozovPRA2012} that treated
 a single two-level system in a bath.

 To evaluate the expectation value of $J_x$, we first calculate $\langle J_{+}(t)\rangle$, where $J_+ = J_x + iJ_y$ is the standard angular momentum ladder operator. Note first that
\begin{equation}
\langle J_{+}(t) \rangle = \sum_{mn} \rho_{mn}(t) (J_{+})_{nm}.
\end{equation}
Using
\begin{equation}
\opav{n}{J_+}{m} = \sqrt{\left(\frac{N}{2} - m\right)\left(\frac{N}{2} + m + 1\right)} \, \delta_{n,m+1}.
\end{equation}
one obtains
\begin{equation}
\av{J_+(t)} = \sum_m \rho_{m,m+1}(t) \sqrt{\left(\frac{N}{2} - m\right)\left(\frac{N}{2} + m + 1\right)}.
\end{equation}
We then have
\begin{equation}
j_x \equiv \frac{2}{N} \av{J_+(t)} = e^{i\omega_0 t} e^{i\Delta(t)t} e^{- t \gamma(t)} X(t) F_x^c(t),
\end{equation}
where
\begin{align}
X(t) &= \frac{2}{N}\sum_m [\rho_S(0)]_{m,m+1} e^{2i\Delta(t)mt} \sqrt{\left(\frac{N}{2} - m\right)\left(\frac{N}{2} + m + 1\right)}, \\
\label{fxc} F_x^c(t) &= \sum_l p_l e^{-2il\Phi(t)}.
\end{align}

For the initial state $\ket{\psi} = e^{-i \frac{\pi}{2} J_y} \ket{\frac{N}{2}}$ prepared by a projective measurement, it is an eigenstate of $J_x$ with eigenvalue $\frac{N}{2}$. Then,
\begin{equation}
[\rho_S(0)]_{m,m+1} = \opav{m}{e^{-i\frac{\pi}{2}J_y}}{N/2} \opav{N/2}{e^{i\frac{\pi}{2}J_y}}{m + 1}.
\end{equation}
Note that $\opav{m}{e^{-i\alpha J_y}}{m'}$ is a Wigner ``d-matrix" element. Using the Wigner formula, we obtain
\begin{equation}
\opav{m}{e^{-i\frac{\pi}{2} J_y}}{N/2} = \frac{1}{2^{\frac{N}{2}}}\sqrt{\binom{N}{\frac{N}{2}+m}}.
\end{equation}
Then,
\begin{equation}
[\rho_S(0)]_{m,m+1} = \frac{1}{2^N} \sqrt{\binom{N}{\frac{N}{2}+m} \binom{N}{\frac{N}{2} + m + 1}},
\end{equation}
from which we have
\begin{align}
X(t) = \frac{2}{N}\sum_m \frac{1}{2^N} e^{2imt\Delta(t)} \sqrt{\binom{N}{\frac{N}{2}+m} \binom{N}{\frac{N}{2}+m+1} \left(\frac{N}{2} - m\right) \left(\frac{N}{2} + m + 1\right)},
\end{align} which is also given in the main text.
The observable $j_x$ is then given by
\begin{equation}
j_x = e^{-t\gamma(t)}\text{Re}\left[ e^{i[\omega_0 + \Delta(t)]t} X(t) F_x^c(t) \right].
\end{equation}
As to $F_x^{c}(t)$ [see Eq.~(\ref{fxc})] that involves the summation over $p_l$, it can be seen that for large $N$, $F_x^c(t) \approx e^{iN\Phi(t)}$.  That is, due to the exponential factors
$ e^{-\beta \omega_0 l}$ and $ e^{\beta l^2 \mathcal{C}}$, only the $l=-N/2$ term makes a dominating contribution if $N\gg 1$.

We now comment on what happens if, instead of a projective measurement, we use a unitary operation. It can be easily shown that
\begin{equation}
[\rho_S(0)]_{mn} \varpropto \sum_l \opav{m}{\Omega}{l} \opav {l}{\Omega^\dagger}{n} e^{-\beta \omega_0 l} e^{\beta l^2 \mathcal{C}}.
\end{equation}
Once again the term with $l = -\frac{N}{2}$ dominates, so that
\begin{equation}
[\rho_S(0)]_{mn} \approx \opav{m}{\Omega}{-N/2} \opav{-N/2}{\Omega^\dagger}{n}
\end{equation}
By setting the unitary operator $\Omega$ to be $e^{i\frac{\pi}{2}J_y}$, we see that our initial state is approximately the same as in the case of state preparation via projective measurement, with $F_x^{c}(t)\approx e^{iN\Phi(t)}$ for large $N$.  As such, for large $N$ the $j_x$ dynamics are very much the same for the above-mentioned two realizations of initial state preparation.



\end{document}